\documentclass[class=preprint]{nature}
\usepackage{graphicx}
\graphicspath{{Figures/}}

\usepackage[usenames,dvipsnames]{color}
\usepackage[colorlinks=true,citecolor=blue,linkcolor=blue,urlcolor=blue]%
{hyperref}
\usepackage[a4paper]{geometry}
\usepackage{fixltx2e}
\usepackage{upgreek}
\usepackage{amsmath}
\usepackage{amssymb}
\usepackage{amsthm}
\usepackage{mathrsfs}
\usepackage{amsbsy}
\usepackage{amstext}
\usepackage{xcolor}
\usepackage{float}
\usepackage{multirow}
\usepackage{wasysym}
\usepackage{mathtools}
\usepackage{subfigure}
\usepackage[version=4]{mhchem}
\usepackage{amsfonts}
\usepackage{dcolumn}
\usepackage{placeins}
\usepackage{float}
\usepackage{caption}
\usepackage[right]{lineno}
\usepackage{fancyhdr}
\usepackage{bbold,bm}
\usepackage{pdfpages}
\usepackage[utf8]{inputenc}
\usepackage[T1]{fontenc}%

\usepackage{setspace}
\usepackage{siunitx}  
\providecommand{\U}[1]{\protect\rule{.1in}{.1in}}
\DeclareCaptionLabelSeparator{vline}{$\boldsymbol{:}$}
\makeatletter
\let\saved@includegraphics\includegraphics
\AtBeginDocument{\let\includegraphics\saved@includegraphics}
\renewenvironment*{figure}{\@float{figure}}{\end@float}
\makeatother

\title{Discovery of magnon self-interaction in a strongly driven antiferromagnet}

\author{David Rohrbach$^{1\dagger*}$, Zhuquan Zhang$^{1\dagger*}$, Takayuki Kurihara$^2$, and Keith A. Nelson$^{1*}$ \\
	\normalsize{$^1$Department of Chemistry, Massachusetts Institute of Technology, Cambridge, Massachusetts, USA, 02139 }\\	
	\normalsize{$^2$Institute for Solid State Physics, The University of Tokyo, Kashiwa, Chiba, Japan, 277-8581} \\ 
	\normalsize{$^{*}$E-mail: kanelson@mit.edu, zhuquan@mit.edu, drohr@mit.edu} \\
	\normalsize{$^\dagger$These authors contributed equally to this work} \\}
\begin{document}

\maketitle

\newpage
\section*{Abstract}
Nonlinear dynamics govern a wide array of natural phenomena and are essential for understanding nonequilibrium behaviors in condensed matter systems. In magnetically ordered materials, magnons—the quanta of spin waves—exhibit intrinsic nonlinearities that are of great interest in fundamental research and practical applications. Despite progress in the nonlinear control of magnon modes in antiferromagnetic materials, the transition from perturbative to non-perturbative regimes of magnon coherences has remained elusive. Here, we explore the nonlinear dynamics of a magnon mode in an antiferromagnet using two-dimensional terahertz spectroscopy with waveguide-enhanced terahertz fields. By driving the magnon mode far from equilibrium, we demonstrate the emergence of high-order magnon coherences and delineate a distinct transition into non-perturbative magnon nonlinearities. This behavior originates from the intrinsic anharmonicity of the magnetic potential and marks a regime dominated by magnon self-interactions at large spin deflection angles. These findings provide fundamental mechanistic insights that might be exploited for ultrafast switching and other advanced magnonic applications.
\newpage

From the turbulent flow of ocean waves\cite{foias2001navier} to the intricacies of neuronal activities in the brain\cite{friston2001book}, nonlinear dynamics encapsulate some of nature's most complex and fascinating phenomena\cite{strogatz2018nonlinear}. In condensed matter systems, while linear response theory effectively describes equilibrium properties, nonlinear responses emerge when materials are driven out of equilibrium. Recently, coherent optical driving of collective excitations has led to the discovery of various nonlinear phenomena including anharmonic excitations of lattice vibrations (phonons)\cite{forst_nonlinear_2011,mankowsky2014nonlinear,kozina2019terahertz}, interacting spin-wave (magnon) coherences in magnets\cite{zhang2024upconversion,zhang2024coupling,zhang2025parametric}, and amplified electronic modes in superconductors such as Higgs and Josephson plasma modes\cite{buzzi2021higgs,luo2023quantum,dienst2013optical,rajasekaran2016parametric,von2022amplification}. These phenomena generally manifest within a regime where nonlinear responses scale perturbatively with the external pump fields. However, when collective modes are driven sufficiently far from equilibrium, their behaviors can deviate markedly from this conventional perturbative framework, exhibiting distinctly non-perturbative characteristics similar to those observed in strong-field physics\cite{kruchinin2018colloquium,schmid2021tunable,zhang2024keldysh}. Although such non-perturbative dynamics—closely related to ultrafast driven responses including structural phase transitions\cite{nova2019metastable,li2019terahertz} and switching of macroscopic orders\cite{schlauderer2019temporal,zhang2024spin}—have attracted considerable interest, the underlying mechanisms and key factors that drive collective modes into this non-perturbative regime remain poorly understood.

In particular, the quanta of spin waves—magnons—are elementary collective excitations in magnetically ordered materials that exhibit intrinsic nonlinearities\cite{zheng2023tutorial}. With increasing driving force, a transition from low-order to high-order magnon nonlinearities is expected, raising fundamental questions about the regime in which the perturbative picture breaks down. Magnon nonlinearities are also pivotal for applications such as switching of magnetic states through coherent spin excursions\cite{zhang2024spin}, creating exotic quantum states like magnon Bose-Einstein condensates\cite{demokritov2006bose}, and developing magnonic logic gates\cite{chumak2015magnon,mehra2024terahertz}. Recent efforts to extend nonlinear magnonics into antiferromagnetism and altermagnetism using tailored terahertz (THz) pulses have led to discoveries such as magnon multiple-harmonic generation\cite{lu2017coherent,zhang2023generation,huang2024extreme}, magnon-magnon mixing\cite{zhang2024coupling}, and magnon conversion\cite{zhang2024upconversion} and parametric amplification\cite{zhang2025parametric}. However, it remains to be established how magnon nonlinearity progresses in strongly driven antiferromagnets with large spin deflections. Realizing this requires (1) intense THz magnetic fields to drive magnon modes far from equilibrium and (2) a sensitive spectroscopic method capable of distinguishing high-order magnon coherences.

To address these challenges, here we have performed measurements of a model antiferromagnetic insulator YFeO$_3$ using two-dimensional (2D) THz spectroscopy with waveguide-enhanced magnetic fields. By directly driving a magnon mode into a highly nonlinear regime and spectrally disentangling its excitation pathways, we reveal high-order magnon coherences up to the 13th order. We experimentally and numerically demonstrate that these high-order nonlinearities herald a transition from perturbative to non-perturbative magnon nonlinearities, originating from the intrinsic anharmonicity of the magnon coordinate. Our findings suggest that magnon self-interactions are the primary nonlinear responses at large spin deflection angles and serve as precursors to the ultrafast switching of antiferromagnetic orders.

\section*{Results}
\begin{figure}[hbtp]
	\centering
    \includegraphics[width=\columnwidth]{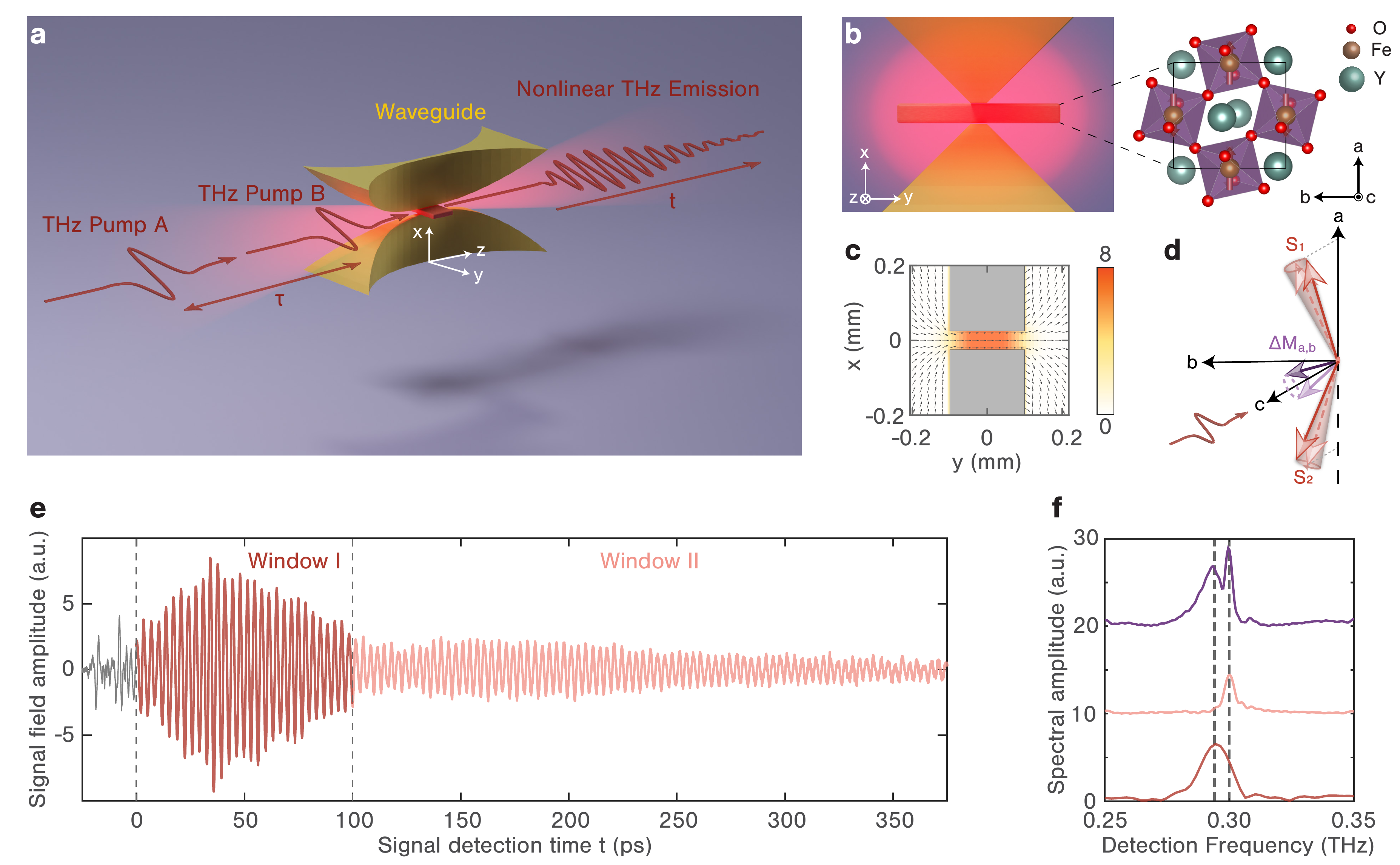}
	\caption{\label{fig:Fig1}
	\textbf{Strong driving of the qFM magnon mode in YFeO$_3$ with enhanced THz fields. a,} Schematic illustration of the experimental setup. A pair of THz pulses is enhanced locally within the waveguide structure, where they interact with the sample to generate nonlinear magnonic emission. \textbf{b,} Left: Cross-sectional depiction of the $a$-cut YFeO$_{3}$ sample positioned inside the waveguide. The sample flake (\SI{50}{\um} thick) is secured by two parallel metallic ridges, defining the minimum gap in the waveguide. Right: Crystal and magnetic structure of YFeO$_3$, an orthorhombically distorted perovskite (space group: Pbnm). Fe$^{3+}$ spins align nearly antiparallel along the $a$ axis, with a slight canting due to the Dzyaloshinskii–Moriya interaction, resulting in a net magnetization $\textbf{M}$ along the $c$ axis. The sample is oriented inside the waveguide such that its $c$ axis aligns with the $z$ direction of the waveguide. \textbf{c,} Simulated THz magnetic field amplitude profile at 0.3~THz in the sample region, illustrating the local field enhancement factor relative to conventional THz free-space focusing. \textbf{d,} Displacement pattern of the qFM magnon mode, characterized by out-of-phase precession of sublattice spins $\mathbf{S}_1$ and $\mathbf{S}_2$ resulting in a precession of the net magnetization $\textbf{M}$. \textbf{e,} Measured nonlinear magnonic emission signal for an inter-pulse delay $\tau$ of 3.4~ps, showing an initial rapid rise and decay (window I, red) followed by a prolonged slow decay (window II, light red). \textbf{f,} Fourier transforms of the nonlinear signal in \textbf{e} for different time intervals: window I (red), window II (light red), and the full window covering both window I and window II (purple).
}
\end{figure}

Figure~\ref{fig:Fig1}a shows our experimental design, wherein a pair of tailored THz pulses with peak free-space magnetic fields around 0.17~T is coupled to a custom-designed metallic waveguide\cite{rohrbach2023wideband} (see Methods and Supplementary Note 1 for more details on the experimental setup). Inside the waveguide, as shown in Fig.~\ref{fig:Fig1}b, a \SI{50}{\um} thick $a$-cut YFeO$_3$ single crystal is sandwiched between parallel metal ridges with its $c$ axis along the THz propagation direction (the $z$-direction). This waveguide structure significantly enhances both the THz electric and magnetic fields, with the simulated magnetic field amplitude spatial profile, illustrated in Fig.~\ref{fig:Fig1}c, showing a field enhancement factor of approximately 8 at 0.3~THz compared to the conventional THz free-space focus. This enhanced magnetic field component is predominantly oriented along the crystallographic $b$ axis of YFeO$_3$ to drive out-of-phase rotations of the two sublattice spins $\mathbf{S}_1$ and $\mathbf{S}_2$ due to magnetic Zeeman interactions (see Fig.~\ref{fig:Fig1}d). The collective motions of the sublattice spins result in an elliptical precession of the net magnetization $\textbf{M}$ around the crystallographic $c$ axis, which corresponds to the quasi-ferromagnetic (qFM) magnon mode.

Using our experimental setup, we first demonstrate that we can drive the qFM mode into a highly nonlinear regime. To achieve this, we set the inter-pulse delay time between the two THz pulses to $\tau = 3.4$~ps and subsequently extract the nonlinear THz emission signals generated by the cooperative interactions of both pulses (see Methods). In stark contrast to all previous 2D THz experiments on magnons\cite{lu2017coherent,zhang2024upconversion,zhang2024coupling,huang2024extreme}, the nonlinear signal here is remarkably strong, reaching a magnitude comparable to that of the free induction decay signal (see Supplementary Note 2A). Figure~\ref{fig:Fig1}e shows the resulting nonlinear response as a function of the signal detection time $t$. Initially, the nonlinear signal builds over time and then rapidly decays within the first 100~ps (0 < $t$ < 100~ps, window I), followed by a much slower decay that extends up to hundreds of picoseconds without vanishing ($t$ > 100~ps, window II). The prolonged decay time indicates the long coherence time of the qFM mode in single-crystalline YFeO$_3$\cite{li2023terahertz}, while the initial increase in amplitude does not align with expected linear responses. Although similar beating patterns may emerge due to magnon-polariton formation\cite{grishunin2018terahertz,sivarajah2019thz,blank2023magneto}, our combined experimental and theoretical analyses exclude that possibility, attributing the observed phenomena solely to the intrinsic nonlinear characteristics of the qFM mode (see Supplementary Note 2B). The anomalous dynamics are further elucidated through frequency domain analysis, as illustrated in Fig.~\ref{fig:Fig1}f. The Fourier transform of the signals from both time windows (I and II) reveals a sharp peak at 0.300~THz, consistent with the qFM mode frequency in linear response (i.e.,  $\Omega_{qFM}=0.300$~THz\cite{yamaguchi2010coherent,jin2013single}), overlaying a broader spectral feature centered at 0.294~THz. The broad peak is reminiscent of the previous observation of coherent spin switching near the spin orientation transition temperature in a similar orthoferrite\cite{schlauderer2019temporal}. In our case, the Fourier spectrum of window II exclusively displays the unshifted qFM mode frequency, while the spectrum from window I reveals the redshifted qFM resonance, associated with the pronounced nonlinear response occurring primarily within the first 100~ps.

These observations motivate us to explore the underlying nonlinear magnon coherences hidden in the 1D spectra. To this end, we record the strong nonlinear signals during the first 100~ps (window I) while varying the inter-pulse delay $\tau$ (from 3 to 103~ps) to construct a 2D time-domain nonlinear response, $\mathbf{H}_{N L}(\tau, t)$, as presented in Fig.~\ref{fig:Fig2}a. These time-domain data clearly show that the nonlinear magnon coherences manifest as oscillatory responses that depend on both $t$ and $\tau$. The oscillations along the $t$ axis indicate the detected frequencies of the nonlinear signals, akin to previous measurements shown in Fig.~\ref{fig:Fig1}e. Meanwhile, the oscillations along the $\tau$ axis encode specific frequency components that modulate the resulting nonlinear signals in and out of phase by altering the inter-pulse delay. By performing a 2D Fourier transformation of the time-domain data, we generate the corresponding 2D THz spectrum $S(\nu,f)$ that correlates the excitation and detection frequencies of the nonlinear magnonic responses. This 2D spectrum provides signatures of nonlinear mode-mode interactions with exceptional clarity\cite{lu2017coherent,maag2016coherent,johnson2019distinguishing,mashkovich2021terahertz,lin2022mapping,blank2023empowering,blank2023two,katsumi2024revealing,leenders2024canted,zhang2024upconversion,zhang2024coupling,zhang2025parametric}. As shown in Fig.~\ref{fig:Fig2}b, we observe a plethora of distinct peaks, all aligning with the detection frequency of the strongly driven qFM mode (i.e., $\Omega_{qFM}^{'}=0.294$~THz), but with different excitation frequencies. A spectral line-cut along $f = \Omega_{qFM}^{'}$, displayed in Fig.~\ref{fig:Fig2}c with a logarithmic amplitude scale, reveals evenly distributed peaks separated by integer multiples of the driven qFM frequency (i.e., $\nu = m \Omega_{qFM}^{'}$, where $m$ is an integer) extending up to $\nu = 7 \Omega_{qFM}^{'}$ and $\nu = -7 \Omega_{qFM}^{'}$ observable above the noise floor. These signals, termed multiple-quantum peaks\cite{turner2010coherent,yu2019observation}, fundamentally differ from previously reported magnon high harmonic generation that produces signals at multiple harmonics of the magnon frequency\cite{zhang2023generation,huang2024extreme}, which would appear as peaks along the detection frequency axis in the 2D spectra. In contrast, the signals we observed result from the initial terahertz pulse interacting multiple times with the magnon mode to generate non-radiative high-order (multiple-quantum) coherences, which are then converted into radiative single-quantum magnon coherences through subsequent interactions with the second terahertz pulse.

To validate that these signals are generated by the enhanced THz magnetic fields instead of the electric fields, we simulated the corresponding time-domain responses and the 2D spectra by solving the Landau–Lifshitz–Gilbert (LLG) equation based on a two-spin Hamiltonian with time-dependent effective magnetic fields (see Methods and Supplementary Note 3 for simulation details). The simulation results, presented in Figs.~\ref{fig:Fig2} d-f, align excellently with the experimental data and replicate key features observed both in time and frequency domains, including the gradual buildup of signals with time $t$ and the emergence of multiple-quantum peaks in the 2D spectrum. 

We further elucidate the evolution of these peaks depending on the THz magnetic field strength. In a perturbative framework, non-rephasing multiple-quantum signals (e.g., peaks at [$\Omega_{qFM}^{'}$, $n\Omega_{qFM}^{'}$], where $n$ is a positive integer greater than 1) involve $2n-1$ field interactions, scaling as a power law with respect to the THz magnetic fields ($\propto \mathbf{H}_{THz}^{2n-1}$) represented by a straight line on a bi-logarithmic scale. This scaling arises because the initial excitation of the $n$-quantum coherence requires $n$ field interactions, while reestablishing the single-quantum coherence in phase with the initial $n$-quantum coherence requires an additional $n-1$ field interactions. Similarly, rephasing multiple-quantum signals (i.e., peaks at [$\Omega_{qFM}^{'}$, $-n\Omega_{qFM}^{'}$], where $n$ is a positive integer) involve the initial excitation with 
$n$ field interactions and another $n+1$ field interactions to produce the single-quantum coherence that oscillates out-of-phase with respect to the initial $n$-quantum coherence, thereby entailing a total of $2n+1$ field interactions that scales as $\propto \mathbf{H}_{THz}^{2n+1}$. Figures~\ref{fig:Fig3}a and \ref{fig:Fig3}b showcase the field dependences of the peak amplitudes for both non-rephasing and rephasing signals, including conventional third-order responses (e.g., peaks at [$\Omega_{qFM}^{'}$, $\Omega_{qFM}^{'}$] and [$\Omega_{qFM}^{'}$, $-\Omega_{qFM}^{'}$]) and high-order multiple-quantum signals (see Supplementary Note 4 for a detailed description of the peak assignments and their origins). Up to approximately 80\% of the maximum field strength, all responses roughly follow the expected behaviors in the perturbative regime. However, saturation effects are observed for relative field strengths exceeding 80\%, at which point the perturbative description breaks down. These trends are comprehensively captured by our numerical simulations based on the LLG equation, which incorporate all nonlinear effects up to arbitrary orders.

Not only do the peak amplitudes saturate, but also the frequencies of the high-order responses shift significantly as the driving fields increase. To illustrate this effect, Fig.~\ref{fig:Fig3}c highlights the 2D spectra of the three-quantum signal at [$\Omega_{qFM}^{'}$, $3\Omega_{qFM}^{'}$] across various THz field strengths, alongside corresponding results from numerical simulations. From both experimental and simulated spectra, it is evident that both the detection and the excitation frequencies of the signal peak undergo a redshift as the field strength increases. The redshift in the detected magnon frequency is consistent with the 1D spectra shown in Fig.~\ref{fig:Fig1}f, while the 2D spectra reveal a more pronounced change manifested along the excitation frequency. The corresponding spectral line cuts also demonstrate that increasing THz field strengths result in a more asymmetric peak profile, with more spectral weight at lower excitation frequencies.

\section*{Discussion}
These features indicate that the magnon mode is driven far from equilibrium, deviating from a perturbative regime. To uncover the origin of the responses, we analyze the semiclassical spin dynamics in more detail. Figure~\ref{fig:Fig4}a displays the three-dimensional trajectories of the two sublattice spins and the net magnetization $\textbf{M}$ resulting from the coherent excitation of the qFM mode. Following THz excitation, the relative canting angle between the two spins remains almost fixed, and the magnetization $\textbf{M}$ deflects from the $c$ axis, tracing a distorted elliptical orbit. Thus, we consider the simplified LLG equation in polar coordinates (see Supplementary Note 5):
\begin{equation}
\label{eqn:Eqn1}
    \frac{\partial^2 \varphi}{\partial^2 t}+\frac{1}{2}\Omega_{qFM}^2 \sin{2\varphi} + 2\zeta \frac{\partial \varphi}{\partial t} = \mathcal{F}(t),
\end{equation}
where $\varphi$ is the deflection angle of $\textbf{M}$ projected onto the crystallographic $ac$ plane, $\zeta$ is a phenomenological damping constant, and $\mathcal{F}(t)$ is the time-dependent driving field. This equation of motion closely resembles that of a classical nonlinear pendulum\cite{strogatz2018nonlinear}. For small $\varphi$, the equation reduces to that of a harmonic oscillator, capturing the excitation of the qFM magnon mode within the linear response regime. However, with significant increases in $\varphi$ under strong external driving fields, the small-angle approximation breaks down, necessitating the inclusion of higher-order terms such as $\varphi^3$, $\varphi^5$, etc. Analogous to a classical nonlinear pendulum swung to large angles, the anharmonic effects at increased deflection angles lead to slowed precession of $\textbf{M}$, effectively softening the qFM mode. Moreover, the nonlinear terms in $\varphi$ reveal that the self-interaction of the qFM mode is non-negligible, leading to the formation of hierarchical magnon nonlinearities up to arbitrary orders. We elucidate these nonlinear dynamics responsible for the observed high-order responses in Fig.~\ref{fig:Fig4}b through a three-step process: (1) $n$ field interactions initiate a hierarchical response involving $n$ self-interacting magnon coherences that are in phase with each other; (2) additional $n-1$ ($n+1$) fields interact with the system, with phases opposite to those of the initial $n$ interactions; (3) following these interactions, the magnon mode resumes oscillation at its fundamental frequency (i.e., one-quantum magnon coherence) and is detected as a radiative photon field. The resultant nonlinear response shares similarities with optical Kerr effects\cite{loriot2009measurement}, primarily resulting in frequency shifts without generating new frequencies. Therefore, the softening of the magnon mode away from equilibrium can be regarded as generalized Kerr-type nonlinear responses\cite{weissl2015kerr,wang2021nonreciprocal,ji2023kerr,zheng2023tutorial} due to magnon self-interactions, where each field interaction resonates with the magnon coherence through the THz magnetic field components.

\begin{figure}[hbtp]
	\centering
    \includegraphics[width=\columnwidth]{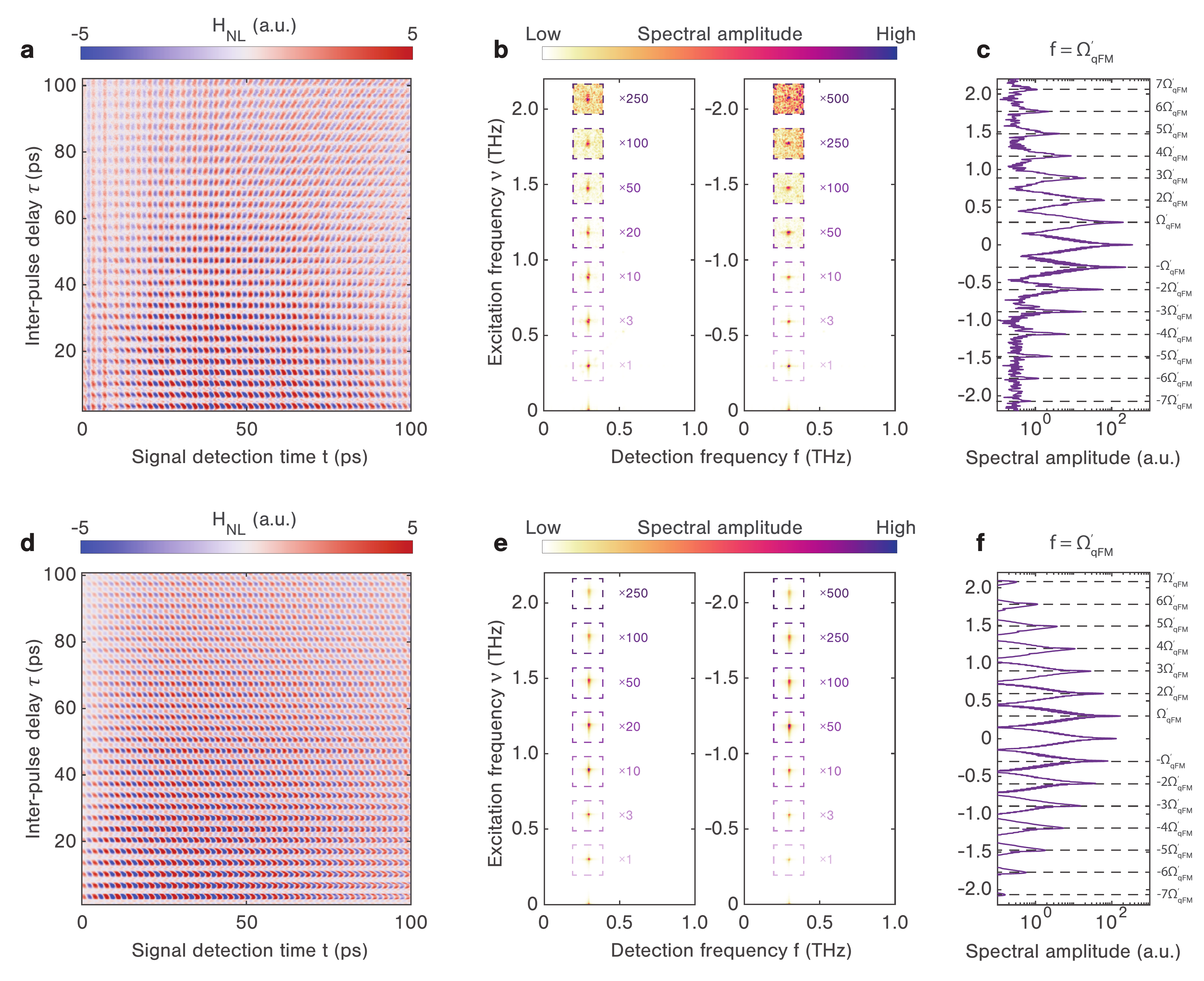}
	\caption{\label{fig:Fig2}  
	\textbf{Experimental and simulated nonlinear 2D THz spectra of the strongly-driven qFM mode. a,} Measured 2D time-time plot of $\mathbf{H}_{N L}(\tau, t)$ from the qFM mode.
    \textbf{b,} 2D THz spectrum of the qFM mode obtained from the 2D Fourier transform of the data in \textbf{a}. For clarity, the non-rephasing (i.e., $\nu > 0$) and the rephasing (i.e., $\nu < 0$) parts of the 2D spectrum are displayed separately on the left and right, respectively. Selected spectral features are scaled by the indicated factors to highlight high-order coherence signals at integer multiples of the driven magnon frequency (i.e., at $\nu = m \Omega_{qFM}^{'}$, where $m$ is an integer).
     \textbf{c,} Spectral line-cut through the 2D spectrum at the detection frequency matching the driven qFM magnon frequency (i.e., $f=\Omega_{qFM}^{'}$).
     \textbf{d-f,} Corresponding numerical simulation results. 
     }
\end{figure}

\begin{figure}[hbtp]
	\centering
    \includegraphics[width=\columnwidth]{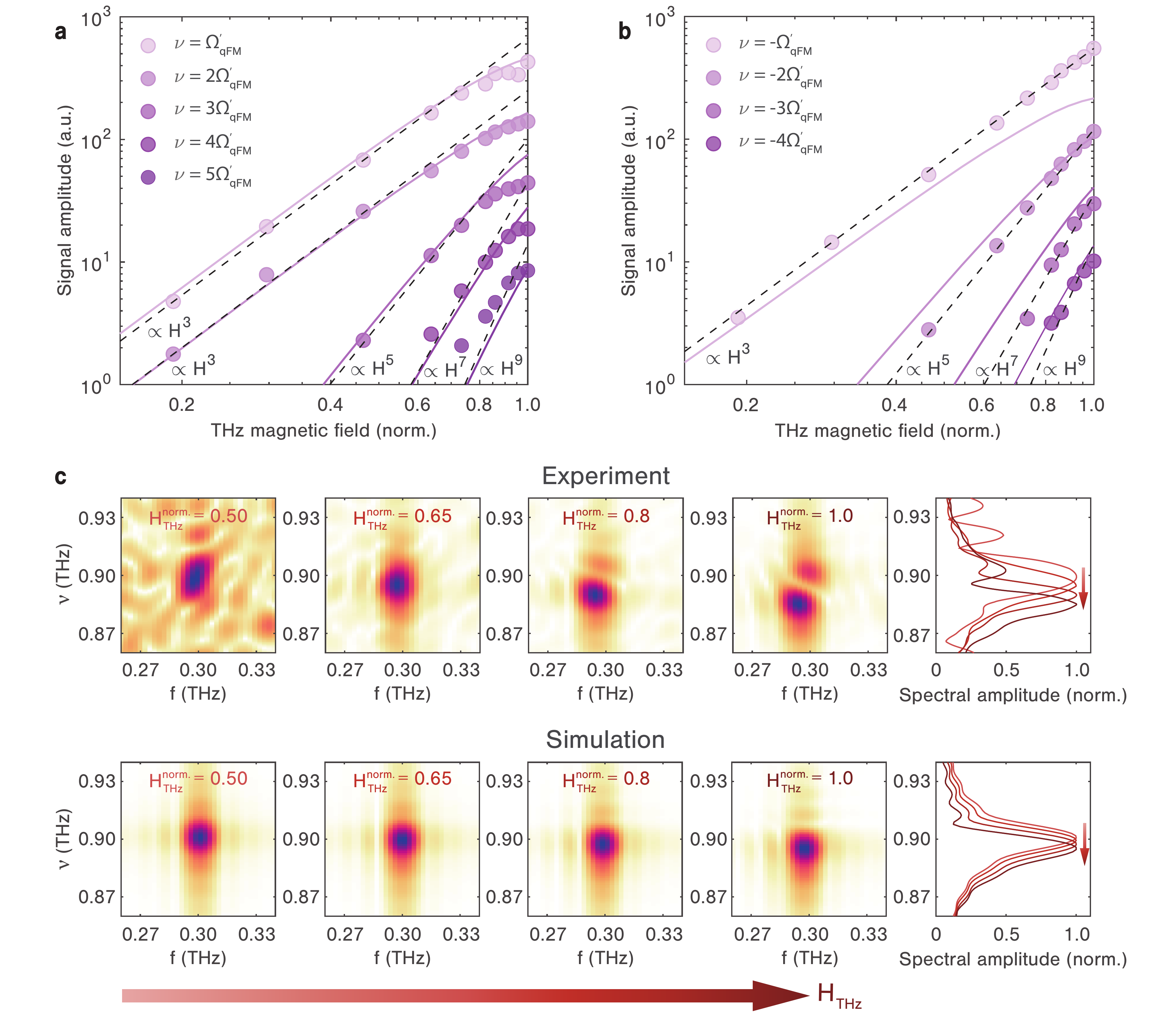}
	\caption{\label{fig:Fig3} 
	\textbf{Field dependence of the high-order coherence signals.} Measured peak amplitudes (circles) as a function of THz magnetic field strength for various non-rephasing (\textbf{a}) and rephasing (\textbf{b}) signals. Solid curves represent results from LLG simulations. Dashed black lines are fits to the experimental data, following the expected power-law dependence in the perturbative regime. Both experimental and simulation results deviate from the power-law dependence at higher THz field strengths.
    \textbf{c,} Selected 2D THz spectra highlighting the peaks with $\nu = 3 \Omega_{qFM}^{'}$ at four distinct field strengths, with experimental measurements shown at the top and numerical simulations at the bottom. The right column shows corresponding spectral line cuts along the excitation frequency at $f = \Omega_{qFM}^{'}$, normalized to their peak amplitudes. Both experimental and simulation results exhibit a redshift in the peak frequency as the THz field strength increases. The more pronounced frequency shift observed in experimental data likely results from inhomogeneous magnetic field profiles, a complexity not fully captured by simulations assuming a uniform THz field strength.
    }
\end{figure}

\begin{figure}[hbtp]
	\centering
     \includegraphics[width=\columnwidth]{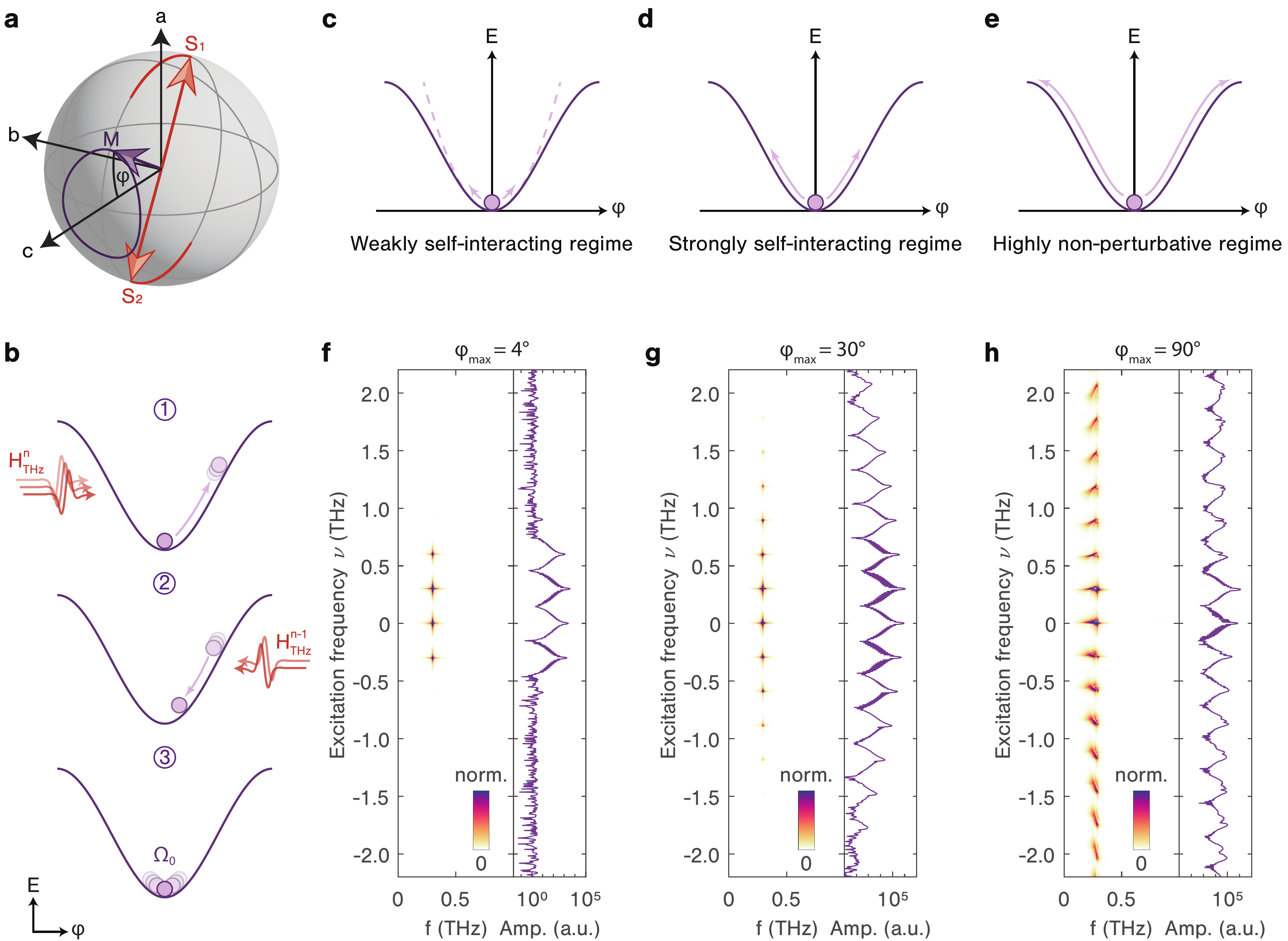}
	\caption{\label{fig:Fig4}  
	\textbf{Origin of the high-order magnon coherences. a,} Bloch sphere representation of trajectories of the two sublattice spins $\mathbf{S}_1$ and $\mathbf{S}_2$ and the magnetization $\textbf{M}$ (rescaled for clarity) when the qFM mode is driven far from equilibrium. $\varphi$ denotes the deflection angle of $\textbf{M}$, defined as half of the central angle of the sector enclosed by $\textbf{M}$ in the $ac$ plane. \textbf{b,} Illustration of magnon displacement driven by sequential THz excitations that result in high-order multiple-quantum responses. The first THz pulse interacts with the magnon mode $n$ times, creating an initial excitation at $n\Omega_{qFM}^{'}$, while the second THz pulse excites the magnon mode $n-1$ times with an opposite phase ($-(n-1)\Omega_{qFM}^{'}$). Together, these interactions causes the mode to oscillate and radiate at $n\Omega_{qFM}^{'} - (n-1)\Omega_{qFM}^{'} = \Omega_{qFM}^{'}$, which corresponds to the peak at [$\Omega_{qFM}^{'}$, $n\Omega_{qFM}^{'}$] in the 2D spectrum. The conjugate process, where the first pulse interacts $n$ times and the second interacts $n+1$ times to produce the single-quantum magnon coherence, generates the peak at [$\Omega_{qFM}^{'}$, $-n\Omega_{qFM}^{'}$]. \textbf{c-e,} Illustrations of magnon displacement on the potential energy landscape, representing weak, strong, and highly non-perturbative self-interacting regimes, respectively. \textbf{f-h,} Simulated 2D THz spectra of the qFM mode obtained by solving the LLG dynamics in polar coordinates, with maximum deflection angles of 4$^{\circ}$, 30$^{\circ}$, and 90$^{\circ}$, respectively. To the right of each 2D spectrum is the corresponding spectral line-cut at the detection frequency matching the driven qFM magnon frequency (i.e., $f=\Omega_{qFM}^{'}$).
}
\end{figure}

In order to delineate the progression of magnon coherences from perturbative to non-perturbative regimes, we numerically solve Equation~\ref{eqn:Eqn1} with $\mathcal{F}(t)$ modeled as two sequential THz pulses (see Methods). We simulate the magnetization dynamics and corresponding 2D spectra for maximum deflection angles ($\varphi_{\text{max}}$) of 4$^{\circ}$, 30$^{\circ}$, and 90$^{\circ}$, representing weak (Fig.~\ref{fig:Fig4}c), strong (Fig.~\ref{fig:Fig4}d), and highly non-perturbative self-interacting (Fig.~\ref{fig:Fig4}e) regimes, respectively. At a small $\varphi_{\text{max}}$ value  of 4$^{\circ}$, the magnon nonlinearity is predominantly governed by third-order self-interacting responses, resulting in four peaks in the 2D spectrum (Fig.~\ref{fig:Fig4}g), consistent with previous experimental observations in the low-order perturbative regime\cite{lu2017coherent,zhang2024coupling}. As $\varphi_{\text{max}}$ increases, more peaks emerge due to the presence of higher-order magnon self-interactions. Indeed, as $\varphi_{\text{max}}$ reaches 30$^{\circ}$, several new peaks appear in the 2D spectrum (Fig.~\ref{fig:Fig4}g), extending along the excitation axis with spectral amplitudes that diminish as the nonlinear order increases. These spectral features align with the data presented in Fig.~\ref{fig:Fig2}. As demonstrated by our field-dependent measurements in Fig.~\ref{fig:Fig3}, within this regime, the spectral amplitudes of various peaks deviate from the perturbative power-law scaling, and peak positions show considerable frequency shifts, marking a transition toward non-perturbative behavior. When $\varphi_{\text{max}}$ is further increased to 90$^{\circ}$, $\textbf{M}$ approaches the switching threshold. As shown in Fig.~\ref{fig:Fig4}h, all peaks in the 2D spectrum stretch significantly toward lower frequencies due to the pronounced softening of the magnon mode. Moreover, compared to the spectrum at $\varphi_{\text{max}}$ = 30$^{\circ}$, high-order peaks continue to intensify while low-order peaks saturate, providing clear spectral fingerprints of the system settling into non-perturbative dynamics.
    
The current study highlights the emergence of high-order magnon coherences arising from the magnon self-interactions in a canted antiferromagnet under intense THz fields. As the antiferromagnetic spins are significantly deflected from equilibrium, we establish the gradual transition from perturbative to non-perturbative dynamics through our combined experimental-theoretical analysis. Experimentally, the implementation of a waveguide THz field enhancement platform extends the capabilities of 2D THz spectroscopy measurements. Future advancements in waveguide geometry and THz field strength optimization may enable each THz pulse to deflect the magnetization by angles exceeding 90$^{\circ}$, thereby functioning as "$\pi/2$" or "$\pi$" pulses akin to those used in coherent control protocols such as Rabi oscillations\cite{brune1996quantum}, Ramsey interferometry\cite{cronin2009optics}, Hahn-echo\cite{hahn1950spin} and nuclear magnetic resonance\cite{hore2015nuclear} measurements. Thus, our experimental approach could open new possibilities for both classical and quantum coherent control of spin states at THz frequencies. More broadly, the access to and detailed characterization of non-perturbative coherences can be generalized to various collective excitations across different material classes, offering new avenues for non-equilibrium control over material states.

\noindent\textbf{Acknowledgments} 
We thank Yu-Che Chien and Tianchuang Luo for their helpful discussions. The work at MIT was supported by the U.S. Department of Energy, Office of Basic Energy Sciences, under Award No. DE-SC0019126. D.R. was funded by the Swiss National Science Foundation (SNSF) as part of a Postdoc Mobility fellowship.

\noindent\textbf{Author contributions}
D.R. and Z.Z. designed the research, performed the experiments, analyzed the experimental data, and performed theoretical analysis and simulations; T.K. provided the sample; D.R., Z.Z., and K.A.N. wrote the manuscript. K.A.N supervised the project.

\noindent\textbf{Competing interests} The authors declare no competing interests.

\newpage

\normalsize
\section*{Methods}
\noindent \textit{Sample Preparation} \\
A single crystal of YFeO$_3$ grown by a floating zone melting technique was used in this work. The crystal was cut perpendicular to the $a$ axis, polished to the thickness of \SI{50}{\um}, and cut to lateral dimensions of about 2~mm by 0.5~mm for installation in the waveguide structure. 

\noindent \textit{2D THz Spectroscopy} \\
The 2D THz spectroscopy setup (Supplementary Fig.~S1a) is powered by the output of a 1~kHz Ti:Sapphire regenerative amplifier system that delivers pulses with 12~mJ energy and 35~fs duration. The majority of the laser pulse energy was split evenly and recombined in a MgO:LiNbO$_3$ (LN) prism with a controlled relative time delay $\tau$. This configuration generated a pair of single-cycle THz pulses through optical rectification using a tilted pulse-front technique\cite{yeh2007generation}. These THz pulses were focused onto the YFeO$_3$ sample mounted within a waveguide structure using two off-axis parabolic mirrors arranged in a $4f$ imaging configuration. To minimize background signals from ambient air, the sample and waveguide assembly were housed inside a custom-built vacuum chamber. The transmitted THz signal was recollimated and refocused onto a 2-mm thick (110) oriented ZnTe crystal by another pair of off-axis parabolic mirrors, and detected via electro-optic (EO) sampling. For THz signal field measurement, a small portion of the laser power was reflected by an echelon mirror, generating an array of 500 spatially shifted pulses with successive time delays of approximately 40~fs, spanning a 20~ps time window for single-shot detection\cite{gao2022high,dastrup2024optical}. The pulse array was focused and overlapped with the THz signals in the EO crystal and imaged onto a high-speed camera that captured the EO sampling traces at the full 1~kHz repetition rate of the laser amplifier. Additional details of our single-shot detection method were discussed elsewhere\cite{gao2022high}.

For 2D THz measurements, we implement a differential chopping scheme where one optical beam pumping the LN crystal (pulse A) was chopped at 250~Hz, and the other (pulse B) at 500~Hz. The time-domain nonlinear signal was extracted from the pulse sequence using the equation:
\begin{equation}
    \mathbf{H}_{NL}(\tau,t) =\mathbf{H}_{A B}(\tau, t)-\mathbf{H}_A(\tau, t)-\mathbf{H}_B(t)+\mathbf{H}_0(t),
\end{equation}
where $\mathbf{H}_{A}$, $\mathbf{H}_{B}$, $\mathbf{H}_{AB}$, and $\mathbf{H}_0(t)$ represent the THz signals measured with pump A, pump B, both pumps, and no pumps, respectively. These signals were collected over 5000 shots (5~s) for each inter-pulse delay $\tau$ and recorded across five different single-shot detection windows along $t$ to cover the entire range from 3~ps to 103~ps. A 2D Fourier transform of the time-domain nonlinear signals along both $t$ and $\tau$ produces the 2D THz spectrum. For field-dependence measurements, a pair of wire grid polarizers was placed before the sample, allowing adjustment of the peak field level. 

\noindent \textit{Waveguide manufacturing} \\
The waveguide assembly consists of two mirror-symmetric parts manufactured from aluminum using standard CNC-machining techniques. Supplementary Fig.~S1b presents a technical drawing of one half of the waveguide structure. After CNC-machining, the contact surfaces of the two components were polished to achieve a surface roughness $R_a$ of less than \SI{0.3}{\um}. A thin YFeO$_3$ crystal was directly installed in the waveguide gap, as shown in the photograph in Supplementary Fig.~S1c.

Coupling the THz pulses to the waveguide mode shrinks the THz mode volume below the free-space diffraction limit and therefore increases the local THz energy density. As a result, both the electric and magnetic fields are enhanced compared to conventional free-space focusing, resulting in a quasi-TEM mode inside the waveguide. The frequency-dependent field enhancement is proportional to the corresponding free-space THz wavelength and increases with smaller gap sizes. The minimum gap size is constrained by the sample thickness, approximately \SI{50}{\um}. In contrast to other field-enhancing structures such as split-ring resonators or spiral microstructures, the waveguide has no intrinsic resonant frequency that requires tuning to the magnon frequency  (see Supplementary Fig.~S2).

\noindent \textit{Numerical simulations} \\
The nonlinear dynamics of magnons induced by THz magnetic fields are numerically simulated based on the following Hamiltonian:
\begin{equation}
\begin{aligned}
\mathcal{H}= & \mathcal{H}_0+\mathcal{H}_{\text {Zeeman }} \\
= & n J \mathbf{S}_1 \cdot \mathbf{S}_2+n \mathbf{D} \cdot\left(\mathbf{S}_1 \times \mathbf{S}_2\right)-\sum_{i=1,2}\left(K_a S_{i a}^2+K_c S_{i c}^2\right) \\
& -\gamma\left[\mathbf{H}_A(\tau, t)+\mathbf{H}_B(t)\right] \cdot\left(\mathbf{S}_1+\mathbf{S}_2\right),
\end{aligned}
\end{equation}
where $J$ is the Heisenberg exchange constant, $n$ is the number of the nearest neighboring spins, $D$ is the antisymmetric exchange constant, $K_a$ and $K_c$ are the onsite magnetic anisotropy constants for the $a$ and $c$ axes, respectively, $\gamma=\frac{g\mu_B}{\hbar}$ is the gyromagnetic ratio, and $\mathbf{H}_{A, B}$ represent the magnetic field components of two time-delayed THz pulses. The parameter values are summarized in Supplementary Table~S1. From a theoretical standpoint, magnon nonlinearities can be systematically treated by the Holstein-Primakoff power expansion of the linear spin-wave Hamiltonian\cite{gyamfi2019introduction}.

From this model Hamiltonian, the time-dependent effective magnetic field is determined as $\mathbf{H}_i^{eff}=-\frac{1}{\gamma}\frac{\partial\mathcal{H}}{\partial S_i}$. Accordingly, the corresponding LLG equation for each sublattice spin $\mathbf{S}_i$ is expressed as:
\begin{equation*} 
	\frac{d\mathbf{S}_i}{dt}=-\frac{\gamma}{1+\alpha^2}[\mathbf{S}_i \times \mathbf{H}_i^{eff}+\frac{\alpha}{|\mathbf{S}_i|}\mathbf{S}_i \times (\mathbf{S}_i \times \mathbf{H}_i^{eff})].
 \end{equation*}
Here, $\alpha$ is a phenomenological damping constant which accounts for energy dissipation. To simulate the 2D time-domain responses, the equations of motion are solved in the presence of both THz pulses, with individual contributions from each THz pulse subtracted to isolate the nonlinear response of the net magnetization $\textbf{M}=\mathbf{S}_1+\mathbf{S}_2$ as a function of both $\tau$ and $t$. A 2D Fourier transformation with respect to $\tau$ and $t$ is then performed to generate the simulated 2D THz spectra.

Additional simulations of the 2D spectra shown in Fig.~\ref{fig:Fig4} are based on the LLG equation in polar coordinates:
\begin{equation}
    \frac{\partial^2 \varphi}{\partial^2 t}+\frac{1}{2}\Omega_{qFM}^2 \sin{2\varphi} + 2\alpha \frac{J S}{\hbar} \frac{\partial \varphi}{\partial t} = -\gamma\frac{\partial}{\partial t}\left[\text{H}_A(\tau, t)+\text{H}_B(t)\right],
\end{equation}
This equation directly connects the nonlinear magnon dynamics with the magnetization deflection angle, $\varphi$. Similar to the above, this equation is solved with each THz pulse present individually and with both pulses together. The nonlinear response of the deflected net magnetization $\textbf{M}\sin{\varphi}$ is calculated, and its 2D Fourier transform with respect to $\tau$ and $t$ generates the simulated 2D THz spectra.

\noindent\textbf{Data availability}
Source data are provided with this paper. All other data that support the findings of this study are available from the corresponding authors upon request.

\noindent\textbf{Code availability}
The codes used to perform the simulations and to analyse the data in this work are available from the corresponding authors upon reasonable request.

\newpage
\section*{References}

\footnotesize
\bibliographystyle{naturemag}

\bibliography{references_new}

\end{document}